\documentclass[]{spie} 
\usepackage[]{graphicx}

\usepackage{caption}
\usepackage{subcaption}
\usepackage{pgfplots}
\usepackage{adjustbox}
\usepackage{multicol}

\title{Botnet Campaign Detection on Twitter} 

\author{Jeremy D. Fields\supit{a,b}
\skiplinehalf
\supit{a}Department of Computer Science, SUNY Polytechnic, Marcy, NY\\
\supit{b}Critical Technologies Inc., 1001 Broad St., Utica, NY\\
}

\authorinfo{Further author information: (Send correspondence to Jeremy D. Fields)\\
Jeremy D. Fields: E-mail: fieldsj@sunyit.edu \\ }
 
\begin{document} 
  \maketitle 

\begin{abstract}
This is an approach to detecting a subset of bots on Twitter, that at best is
under-researched. This approach will be generic enough to be adaptable to most, if not all
social networks. The subset of bots this focuses on are those that can evade most, if not all
current detection methods. This is simply because they have little to no information
associated with them that can be analyzed to make a determination. Although any account
on any social media site inherently has information associated with it, it is very easy to blend in with the
majority of users who are simply “lurkers” - those who only consume content, but do not
contribute. How can you determine if an account is a bot if they don’t do anything? By the
time they act, it will be too late to detect. The only solution would be a real time, or near
real-time, detection algorithm 
\end{abstract}

\keywords{These, Key, Words, Are, Determined, By, The Submission System}

\section{INTRODUCTION}
\label{sec:intro}  
Twitter is a ‘microblogging’ social media website. It stands out from its competitors, such as Facebook and LinkedIn by the fact that it limits posts or tweets (text-based message) to only 140 characters. Twitter is also unique in that relationships can be directed, whereas on sites such as Facebook most relationships are bi-directional. This is made possible due to the way that Twitter allows relationships between users to be created. Twitter is one of the largest websites in the world, and as of the time of this writing, it is ranked the 10th most popular site globally, as reported by Alexa [1]. Similarly, it is ranked as the 8th most popular site in the United States. Twitter boasts having 320 million monthly active users, and over 1 billion unique monthly visits. Furthermore, the company claims that an astounding 500 million tweets are sent every day [2]. 

In this paper we present a novel approach to detecting bots on twitter in near real-time. Our approach comprises of computationally simple comparisons and calculations, as opposed to the all too common machine learning approach to this problem, or non real-time approaches that involve network analysis which is expensive and time consuming.

The subset of bots this method focuses on are those that can evade most, if not all current detection methods. This is simply because they have little to no information associated with them that can be analyzed to make a determination on whether they are a bot or not. While any account on Twitter has inherently has some information associated with it, it is very easy to blend in with the masses of users who are simply ``lurkers'', those who only consume content but do not contribute. How can you determine if an account is a bot or not, especially when they don’t do anything? By the time they act, it's too late to detect them. The only solution would be a real-time, or near real-time, detection algorithm.
 
As stated in previous research, bots can influence public opinion [3,4,5,6,7], especially the reporting done in [7] where the Syrian Intelligence Agency is alleged to have used Twitter and Twitter bots to attempt to shift public opinion. This is certainly an extremely powerful tool, and as with most powerful tools, there is the possibility that it will be used for malicious or less ethical purposes at some point.

While bot detection in general is a highly researched area, detecting large amounts of bots acting in unison and/or in real-time is not. The few works that we could find take non a real-time approach, and rely on other information such as URL analysis and network analysis [8,9].

\section{APPROACH}
This approach, as most others, looks at analyzing simple attributes. The word ‘simple’ is used not so much in reference to the attribute, but the comparisons made upon them. While most researchers have utilized machine learning algorithms such as neural networks and SVMs, the attributes used in this papers approach are analyzed by performing a simple comparison to determine whether they are the same as that tweets neighbors’ attributes. What makes it stand apart is not the attributes used, or even the specific combination, but rather that this approach does not analyze the data, history, or network of an account. Instead, it compares most attributes against the nearest N tweets as they appear in chronological order. This approach has both ups and downs. It has the advantage of only requiring the data of N tweets to make a determination, as the only data considered is that of the nearest N tweets on a chronological scale. As this
approach is designed for detecting bot-nets and spam campaigns, it will not detect a lone bot acting on its own. This approach has massive advantages by requiring so little data that it is non-computationally difficult. This also allows this program to analyze tweets in real-time. 

The determination of whether a user is a bot or not is done by a scoring system. For each attribute below, with the exception of entropy and sentiment, the score for each is on a scale of $0$ to $N$, where $N$ is the number of neighboring tweets that are compared against. As there are 12 attributes, the maximum score is $N * 12$ (with the exception of multipliers).  If a bot meets a certain percentage of this, they are classified as a bot.

\begin{multicols}{2}
\begin{itemize}
 \item Text similarity
 \item For each similar text, how many were within N units of time
 \item Summation of similarity score between current tweet and N neighbors
 \item Language
\item Gender
\item User Agent
\item Time zone
\item Location
\item Profile URL
\item Profile description
\item Entropy
\item Sentiment
\end{itemize}
\end{multicols}

\subsection{Thresholds}
Not all of these attributes are Boolean comparisons of ``are they the same?'', so we must have thresholds for those attributes. Furthermore, not all thresholds or multipliers can be backed by statistical proof, as they are inherently subjective.  We can only try to justify our choice as best we can. Of course, no threshold or multiplier is set in stone and each is easily configurable. They very well may need changing based on data source, or dataset.

\newpage

\subsection{Text Similarity}
For example, how similar to two pieces of text need to be to be considered similar?
\begin{center}
	\begin{adjustbox}{max width=\textwidth}
	\begin{tabular}{ | l | l | l |}
    \hline
    String 1 & String 2 & Score \\ \hline
    
    This is a test & This is a test & 1.0 \\ \hline
    
    abcd & efgh & 0.0 \\ \hline
    
    This is a test & This is another test & 0.83 \\ \hline
    
    (1) Please check out my awesome link! t.co/50938 & (2) Please check out my awesome link!
	t.co/93452 & 0.916 \\ \hline
	
    How different do two strings & have to be to be different? & 0.392 \\ \hline
    
    Apples are good! \#VoteForApple & Oranges are bad! \#VoteForApple & 0.76 \\ \hline

	Apples are good! http://t.co/kjh3sf \#VoteForApple 1 & Oranges are bad! http://t.co/pl5ibq
	\#VoteForApple 2 & 0.725 \\ \hline

    These two strings should & be very different from each other. & 0.103 \\ \hline
    
    Apples are good! http://t.co/kjh3sf \#VoteForApple \#ApplesRock 1 & Oranges are bad! 			http://t.co/pl5ibq \#VoteForApple \#OrangesSuck 2 & 0.677 \\ \hline
    \end{tabular}
    \end{adjustbox}
\end{center}
For this metric, a threshold of 0.6 - 0.65 seems ideal, and thus 0.6 was chosen in this effort.

\subsection{Time Difference, Sentiment, Entropy}

Time difference is the most subjective. A value of 4,000ms was chosen only based on that it should be large enough to account for a bot master to send instructions to all, or most bots, such that most will act within this time frame. As we analyze a sliding window based on a chronological scale, only some of them need to act at any given time.

Standing on the shoulders of others, sentiment \& entropy have already previously been shown to have a high correlation with bots [10,11,12].

Both the sentiment and entropy metrics have been reproduced in our effort independently. Our results were exactly in line with the works of others, which further re-enforces their validity.

\subsection{Neighbors compared against}
Counter-intuitively, the fewer neighbors compared, the higher the number of bots detected. However, if it's too low, then we give up some accuracy. Based off of the tables below, we chose N to be 20. Also, as N grows, so does the time taken to analyze. Fortunately, a low N is desired which helps make this algorithm fast.

This work (and all work in this effort) was done on a machine with a consumer grade Intel Core i7 3770k. This particular example was on a data-set that contained approximately 10,000 tweets.

\begin{center}
	\begin{adjustbox}{max width=\textwidth}
	\begin{tabular}{ | l | l | l | l |}
    \hline
    \# of neighbors compared & Tweets per second(analyzed) & Time taken & Bots detected \\ \hline
    10 & 65.6556 & 2m 31s & 130 \\ \hline
    20 & 34.3044 & 4m 49s & 125 \\ \hline
    30 & 24.299 & 6m 48s & 119 \\ \hline
    40 & 23.4761 & 7m 52s & 115 \\ \hline
    50 & 18.6037 & 9m 42s & 111 \\ \hline
    \end{tabular}
    \end{adjustbox}
\end{center}

This analysis was performed on several data-sets to determine the best value of N. For the sake of brevity, only this partial table is included. In all cases, as N grows, bots detected will start to fall dramatically. In this data-set, when N=100, only 45 bots were detected. However, in all cases, when N=10, although more were detected, it included some false positives. For this reason, N is chosen to be 20.
\newpage
\subsection{High score}

To choose a high score threshold, we analyzed the scores of multiple data-sets. As expected, only a small percentage have an abnormal high score, but it's clear that this spike is where the bots are.

All data-sets looked similar to this graph, but varied on at what percentage the spike occurred. All data-sets fell within a window of ~0.18-0.32.

\begin{figure}[!h]
  \centering
  \includegraphics[scale=0.85]{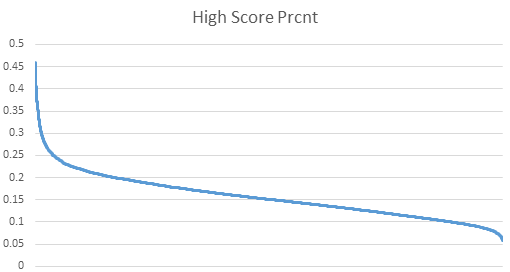}
  \caption{High Score Curve - Count vs High Score}
  \label{fig:superbowl1_hs_curve}
\end{figure}

To get a better feel for what our high score threshold should be, we can also look at the same data through the lens of k-means clustering
\\

\begin{figure}[!h]
  \centering
  \includegraphics[scale=0.45]{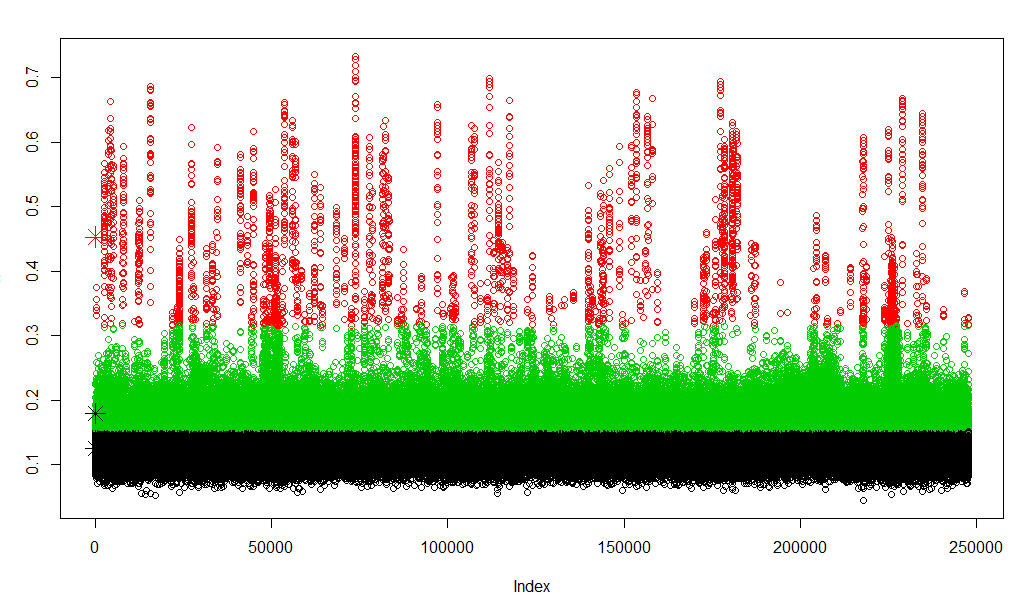}
  \caption{K-Means clustering on high score - 250k tweets}
  \label{fig:politics_250k_hs_cluster}
\end{figure}

Note that this graph is of 250k tweets on a chronological scale. We can see that the spikes comprise only a small percentage of this, but this graph makes them clear.

The threshold certainly changes from data-set to data-set, but overall, 0.25 is a good middle ground. For this reason, 0.25 is chosen as our threshold.

\subsection{Multipliers}
Not all attributes are equal. As shown by others [10,11,12,14,15,16] (and many more), some attributes have a high correlation with an account being a bot, while other attributes are used for reinforcement.

Intuitively, if two tweets have similar, or the same text within the sliding comparison window on a chronological scale (which typically covers only a few milliseconds), it is a very good indicator that those messages were from a bot, and thus we give that individual score a higher multiplier.

\begin{center}
	\begin{adjustbox}{max width=\textwidth}
	\begin{tabular}{ | l | l |}
    \hline
    Attribute & Multiplier \\ \hline
    Similarity & 2 \\ \hline
    Sentiment & 1.2 \\ \hline
    Entropy & 1.2  \\ \hline
    Similarity Count & 1.2 \\ \hline
    Location & 1 \\ \hline
    Language & 1 \\ \hline
    User Agent & 1 \\ \hline
    Time Zone & 1 \\ \hline
    Profile Description & 1 \\ \hline
    Profile URL & 1 \\ \hline
    Time Difference & 1 \\ \hline
    Gender & 1 \\ \hline
    \end{tabular}
    \end{adjustbox}
\end{center}

\subsection{Example Scoring}

To better illustrate how this method works, below will detail an example walk through of how a tweet's bot score is calculated.

In the neighboring 20 tweets... (N = 20)
\begin{center}
	\begin{adjustbox}{max width=\textwidth}
	\begin{tabular}{ | l | l |}
    \hline
    Attribute & Score \\ \hline
    How many have text similarity greater than similarity threshold? & 17 \\ \hline
    \  -   For each of the above, how many were within 4 seconds?  & 14 \\ \hline
    Summation of similarity between tweets in window? & 14.9582  \\ \hline
    Same language & 15 \\ \hline
    Same Gender & 7 \\ \hline
    Same User Agent & 16 \\ \hline
    Same Time Zone & 8 \\ \hline
    Same Location & 0 \\ \hline
    Same profile URL & 17 \\ \hline
    Same profile description & 16 \\ \hline
    Is entropy lower than entropy threshold? & Yes \\ \hline
    Is sentiment higher than sentiment threshold? & No \\ \hline
    \end{tabular}
    \end{adjustbox}
\end{center}

Now we take those scores, and add them all up while respecting multipliers.

\noindent Score = (17 * 2) + 14 + (14.9582 * 1.2) + 15 + 7 + 16 + 8 + 0 + 17 + 16 + (20 * 1.2) + 0 
\ = 168.94984 \\

\noindent Then we calculate the highest possible score \\
\noindent Total Possible Score  = (20 * 2) + ((20*1.2) * 3) + (20 * 8) \\
\ = 272 \\

\noindent Tweet's score as a percentage = 168.94984/272 \\
\ = 0.621 \\

Lastly, we compare that to our high score threshold of 0.25. Since 0.621 is greater than 0.25, this tweet would be labeled as coming from a bot that's part of a bot-net.

In this example, we can see that 17 tweets in the sliding window of 20 tweets were above the similarity threshold, which is a strong indicator of bot activity. It's possible that these are re-tweets, or it may be the case that many people are tweeting the same thing for whatever reason.

We also see that the summation of the similarity scores between the target tweet and it's neighbors is 14.9582. This is very high compared to the 17 tweets that were above the similarity threshold. If it was 17, that would indicate that all 17 tweets were exactly the same. Since this is a little lower, this would indicate that the bots messages were either randomized to some degree (bots often use an increment number - either as an Arabic numeral, or spelled out) or the messages included a link, which Twitter automatically assigns a unique shortened URL to.

Furthermore, we see that these likely bots all have the same profile URL and profile description, which is common with bots who are trying to spam malicious links.

Lastly, and one of the strongest parts of this method, is that no single attribute will make a determination. We see that 0 locations were the same within these tweets. This likely means that location reporting was turned off, which is often the case.

\section{Data}
Some of the data-sets used are detailed below.

\begin{center}
	\begin{adjustbox}{max width=\textwidth}
	\begin{tabular}{ | l | l | l |}
    \hline
    Name & Number of Tweets & Period \\ \hline
    20k & 13,204 & 26m 43s \\ \hline
    1k & 1,000 & 51s \\ \hline
    jadine & 1,832 & 13h 50m 27s \\ \hline
    isis & 11,038 & 22m 22s \\ \hline
    politics\_250k & 249,999 & 4h 45m 15s \\ \hline
    red4tum & 3,536 & 25m 1s \\ \hline
    sample1 & 52,741 & 18m 59s \\ \hline
	superbowl1 & 13,840 & 33m 56s \\ \hline
    superbowllive & 59,804 & 5m 15s \\ \hline
    2\_clean\_25k & 25,000 & 29m 11s \\ \hline
    ny\_primary\_250k\_1 & 248,618 & 3h 46m 21s \\ \hline
    ny\_primary\_250k\_2 & 248,497 & 3h 47m 36s \\ \hline
    snl & 4,113 & 32m 9s \\ \hline
    \end{tabular}
    \end{adjustbox}
\end{center}

We chose and collected data-sets to cover several scenarios. Some data-sets are filtered, some aren't. Some were collected over a year apart. Some were filtered by language, some weren't. All data is as it came from Twitter's streaming API endpoint, which has tweets in chronological scale. The only modifications to these data-sets was only to remove any bad or malformed tweets, which was only a negligible amount, if any at all per data-set.
\newpage
\section{Results}
The below graph is an example of what an organic network looks like. This is
included for reference, and as an example of a ground truth for comparison against the network analysis of bot-nets. We can see that there are many distinct groups, some large, some small, some interconnected, some aren't.

\begin{figure}[h]
  \centering
  \includegraphics[scale=0.40]{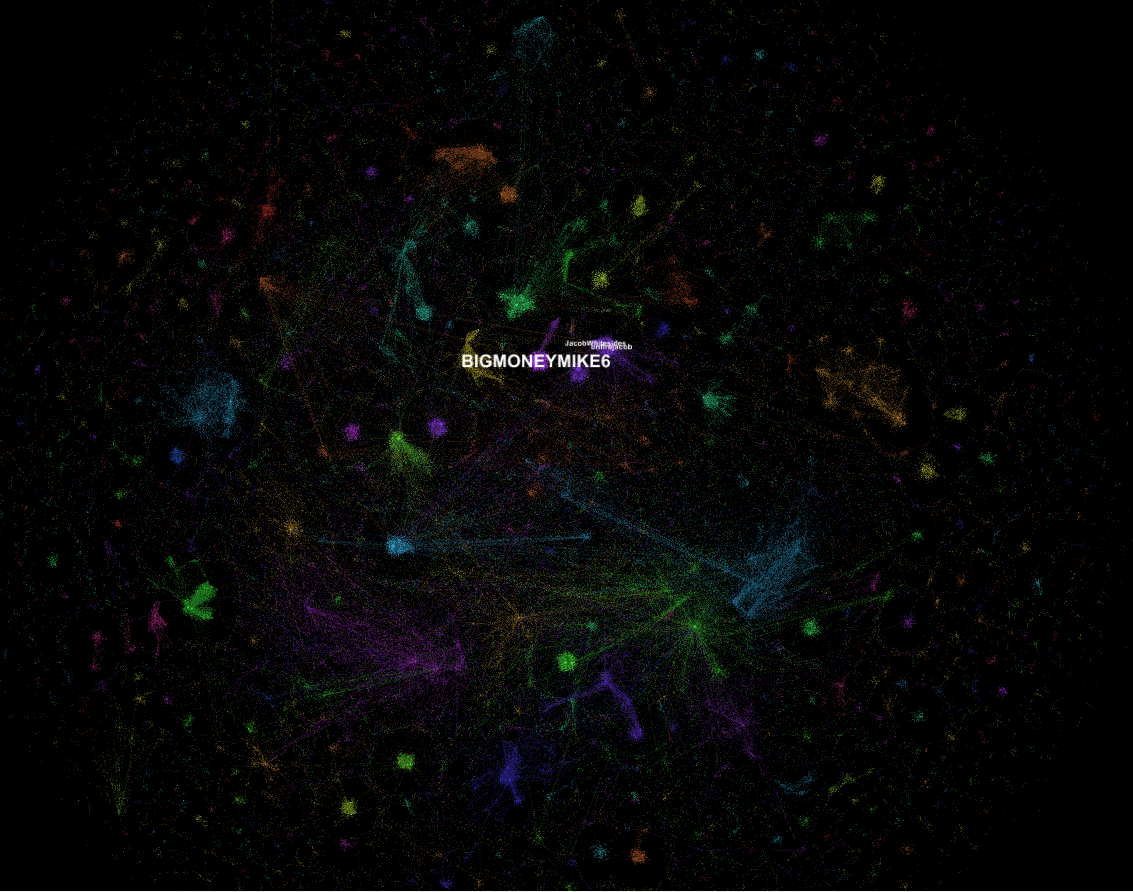}
  \caption{Post analysis of a bot's friends/followers}
  \label{fig:1k}
\end{figure}

Below is a network graph of a detected bot-net. As can be seen, it is one large inter-connected group, where everyone is 'friends' with everyone else, which is highly unusually in an organic network.

\begin{figure}[h]
  \centering
  \includegraphics[scale=0.45]{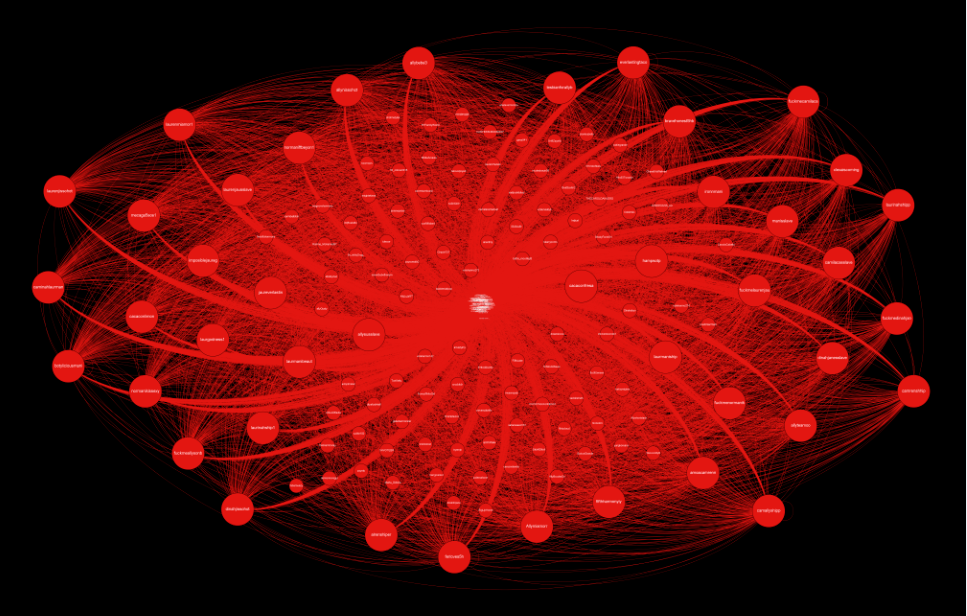}
  \caption{Post analysis of a bot-nets network}
  \label{fig:1k}
\end{figure}

\clearpage

\section{CONCLUSION}
Standing on the shoulders of others, we've shown an approach to detect bot-net spamming campaigns, that is also effective, quick, and can be performed in real-time.

We detected 14,585 bots out of an analyzed 920,008 tweets. Just under 11,000 of these were unique accounts. At the time of this writing, 2,586 of these accounts have been suspended or deleted, which further re-enforces our work. However, it also shows that Twitter is behind on detecting this particular type of bot.

The time coverage for all data-sets collected is 1 day, 5 hours, 33 minutes, and 55 seconds. It took only 5 hours and 45 minutes to analyze this data. Our method was able to analyze data over five times faster than it could be collected. However, none of our data-sets were from Twitter's Firehose stream (100\% of all tweets), and our approach would not work in real time with that amount of data. We're confident that with optimization (i.e. moving to a faster language like C, versus currently using Python), and using commercial grade high-end hardware, our approach would come close, if not achieve in analyzing 100\% of all tweets in real-time. A worst case scenario would simply employ a distributed processing approach.

\section{FUTURE WORK}
There is much to be done with refinement. All multipliers need further research and analysis to determine the best values. The thresholds also need further research into determine with statistical analysis (if possible) what thresholds are ideal.

Also, the attributes used should be further researched. Are all attributes required? Can the same accuracy be obtained with less attributes? Could we achieve higher accuracy with more? What's the minimum amount of attributes required to retain a high accuracy? For example, user agent analysis has been shown to be an indicator of an account being a bot, but this was not something explored in this effort [13].

No optimization were made to our approach, and should certainly be looked into. While this method is quite fast, even using a slow language (Python) on a consumer grade, single PC, it could be much faster.


\newpage
\begin{thebibliography}{9}
\bibitem{alexa} 
\textit{Competitive Intelligence.}. 
Twitter.com Site Overview. Alexa, n.d.
 
\bibitem{twitterblog} 
\textit{"\#numbers | Twitter Blogs}.
"\#numbers | Twitter Blogs. (n.d.) Available: https://blog.twitter.com/2011/numbers
 
\bibitem{zhang}
J. Zhang, R.zhang, Y.Zhang, G.Yan
\textit{On the Impact of Social Botnets for
Spam Distribution and Digital-influence Manipulation}
2013 IEEE Conference on Communications and Network Security (CNS), 2013

\bibitem{lol}
V. S. Subrahmanian, Amos Azaria, Skylar Durst, Vadim Kagan, Aram Galstyan,
Kristina Lerman, Linhong Zhu, Emilio Ferrara, Alessandro Flammini, Filippo Menczer, 
Rand Waltzman, Andrew Stevens, Alexander Dekhtyar, Shuyang Gao, Tad Hogg,
Farshad Kooti, Yan Liu, Onur Varol, Prashant Shiralkar, Vinod Vydiswaran, Quiaozhu
Mei, and Tim Huang.
\textit{The DARPA TWITTER BOT CHALLENGE.}
The DARPA TWITTER BOT CHALLENGE 1 (n.d.): n. pag. ArXiv. DARPA. Available.
https://arxiv.org/ftp/arxiv/papers/1601/1601.05140.pdf.

\bibitem{freitas}
C. Freitas, Fabricio Benevenuto, Saptarshi Ghosh, and Adriano Veloso.
\textit{Reverse Engineering Socialbot Infiltration Strategies in Twitter}
Proceedings of the 2015 IEEE/ACM International Conference on Advances in Social Networks Analysis and Mining 2015 - ASONAM '15: n. pag. ArXiv. 2015.

\bibitem{abokhodair}
 N. Abokhodair
 \textit{Architecture for Understanding the Automated Imaginary: A
Working Qualitative Methodology for Research on Political Bots}
Political Bots. N.p., 09 Aug. 2015.

\bibitem{york}
J.C. York
\textit{Syria's Twitter Spambots}
(T. Guardian, Producer) Retrieved June 2014, Available:
http://www.theguardian.com/commentisfree/2011/apr/21/syria-twitter-spambots-prorevolution

\bibitem{xianchao}
Zhang, Xianchao, Shaoping Zhu, and Wenxin Liang
\textit{Detecting Spam and Promoting Campaigns in the Twitter Social Network.}
2012 IEEE 12th International Conference on Data Mining (2012). Print.

\bibitem{chu}
Z. Chu, I. Widjaja, and H. Wang
\textit{Detecting Social Spam Campaigns on Twitter}
Applied Cryptography and Network Security Lecture Notes in Computer Science: 455-72.
2012.

\bibitem{gianvecchio}
 Z. Chu, S. Gianvecchio, H. Wang, and S. Jajodia
 \textit{Detecting Automation of Twitter Accounts: Are You a Human, Bot, or Cyborg?}
 IEEE Transactions on Dependable and Secure Computing IEEE Trans. Dependable and Secure Comput. 9.6: 811-24, 2012.

\bibitem{white}
J. S. White, and J. N. Matthews
\textit{It's You on Photo?: Automatic Detection of Twitter Accounts Infected with the Blackhole Exploit Kit}
2013 8th International Conference on Malicious and Unwanted Software: "The Americas" (MALWARE), 2013.

\bibitem{hu}
X. Hu, J. Tang, H. Gao, and H. Liu
\textit{"Social Spammer Detection with Sentiment Information.}
2014 IEEE International Conference on Data Mining, 2014.

\bibitem{paxson}
C. M. Zhang and V. Paxson, N. Spring and G. F. Riley (Eds.)
\textit{Detecting and analyzing automated activity on twitter}
 In Proceedings of the 12th international conference on Passive and active measurement (PAM’11), Springer-Verlag, Berlin, Heidelberg, 102-111, 2012.

\bibitem{stringhini}
G. Stringhini, C. Kruegel, and G. Vigna. 
\textit{Detecting Spammers on Social Networks}
Proceedings of the 26th Annual Computer Security Applications Conference on - ACSAC
'10, 2010.

\bibitem{ferrara}
E. Ferrara, O. Varol, C. Davis, F. Menczer, and A. Flammini.
\textit{The Rise of Social Bots}
CoRR(2014): n. pag. ArXiv. Http://arxiv.org/abs/1407.5225, 9 July 2015.

\bibitem{lee}
K. Lee, J. Caverlee, and S. Webb
\textit{Uncovering Social Spammers}
Proceeding of the 33rd International ACM SIGIR Conference on Research and Development in Information Retrieval - SIGIR '10, 2010



\end{thebibliography}
\end{document}